\RequirePackage{ifpdf}
\pdfoutput=1
 \documentclass{JINST}

\usepackage{verbatim}
\newcommand{\MDnewtext}[1]{ #1 }
\newcommand{\margincomment}[1]{\typeout{#1}}

\newcommand{\JMnewtext}[1]{ #1 }

\title{A Radio-Frequency-over-Fiber link for large-array radio astronomy applications.}

\author{Juan Mena\thanks{Corresponding author.},
 Kevin Bandura, Jean-Fran\c{c}ois Cliche, 
 Matt Dobbs, Adam Gilbert, Qing Yang Tang\\
Department of Physics, McGill University\\
   3600 University Street, Montreal, Quebec H3A 2T8, Canada\\
  E-mail: \email{juan.menaparra@mail.mcgill.ca}}

\abstract{A prototype 425-850 MHz
  Radio-Frequency-over-Fiber (RFoF) link for the Canadian Hydrogen
  Intensity Mapping Experiment (CHIME) is presented. The design is
  based on a directly modulated Fabry-Perot (FP) laser, operating at
  ambient temperature, and a single-mode fiber. The dynamic performance,
  gain stability, and phase stability of the RFoF link are
  characterized. Tests on a two-element
  interferometer built at the Dominion Radio Astrophysical Observatory
  for CHIME prototyping demonstrate that RFoF can be successfully used
  as a cost-effective solution for analog signal transport on the
  CHIME telescope and other large-array radio astronomy applications.}

\begin{document}

\section{Introduction}

\margincomment{Juan changed the keywords above since they cannot be
arbitrary but have to be chosen from a list.$\checkmark$}The Canadian Hydrogen Intensity Mapping Experiment 
(CHIME)\footnote{\href{http://chime.phas.ubc.ca/}{www.CHIMExperiment.ca}} \margincomment{this link goes to www.mcgillcosmology.ca/chime $\checkmark$}
is currently under development at the Dominion Radio Astrophysical
Observatory (DRAO) in Penticton, Canada. It is an array of cylindrical
telescopes with \JMnewtext{more than} 2500 radio receivers and has no moving parts or
cryogenics. CHIME will map the three-dimensional distribution of
neutral hydrogen gas in the universe by directly detecting its
redshifted 21-cm radiation. By measuring the scale of the Baryon
Acoustic Oscillations across the redshift range
$\mathrm{z}\approx 0.8$ to $\mathrm{z}\approx 2.5$ in both the angular
and line-of-sight directions, CHIME will study the epoch when Dark
Energy generated the transition from decelerated to accelerated
expansion of the universe.

As a prototype for testing the CHIME technology, a two-element radio
interferometer was built at DRAO. While characterizing the CHIME
analog receiver in the two-element interferometer, the implementation
of a Radio-Frequency-over-Fiber (RFoF) link to transport the \JMnewtext{analog} signals
from the antennas to the processing room has been investigated as an
alternative to traditional coaxial cable. 
Radio astronomy applications of RFoF are becoming common due, in part, to the pioneering implementation \cite{ackcoxdre04}
for the Allen Telescope Array (ATA) \cite{welchetal09}.
RFoF is an attractive option
for the analog receiver in CHIME, and its advantages over coaxial
cable include not only cost, but also small size and light weight,
immunity to electromagnetic interference, low loss, and a separate grounding
scheme for the front end radio frequency (RF) chain and the digital back end.

\JMnewtext{Analog transmission over a RFoF link also reduces the likelihood
of self-generated radio frequency interference (RFI) pickup and reduces the amount of
electronics and maintenance at the antennas compared to a receiver architecture
with the digitizers located at the antennas and digital data transmission to the processing room.}However, 
the \JMnewtext{analog} optical link also introduces challenges that
must be overcome, mainly related to linearity, noise, and stability. In
this paper, a prototype version of a custom-built RFoF link for CHIME
is presented, its dynamic performance, gain stability, and
phase stability are characterized.
The third order intercept point (IP3) is used as the
figure of merit for the RFoF linearity and is appropriate for an
octave-bandwidth system like CHIME. The corresponding spurious-free
dynamic range (SFDR) is used as measure of the RFoF dynamic
performance.\MDnewtext{We demonstrate a system wherein the noise and linearity
are dominated by the 8-bit analog to digital converter (ADC) that follows the
RFoF link, meaning that these challenges have been addressed at level that
makes them negligible for the overall system.}
% The primary design constraint is to have the final
% 8-bit analog to digital converter (ADC) drive the overall system
% performance. 

This paper is structured as follows: Section~\ref{RFoF_req} describes
the present implementation of the analog receiver for the two-element 
interferometer and the
requirements on the dynamic performance of the RFoF link.
Section~\ref{RFoF_perf} describes the analog optical link design and
lab tests on its dynamic performance. In Section~\ref{RFoF_stab}, test
results on the gain and phase stability of the optical link with
changes in temperature of the RFoF transmitter and the optical fiber
are presented. The first tests of the RFoF link in the two-element
interferometer are shown in Section~\ref{RFoF_test}. A final
discussion on the overall RFoF performance and improvements for the
next version of the link are presented in Section~\ref{RFoF_conc}.

\section{Analog signal chain and RFoF performance requirements}
\label{RFoF_req}

\begin{figure} \begin{center}
    \includegraphics[width=1\textwidth]{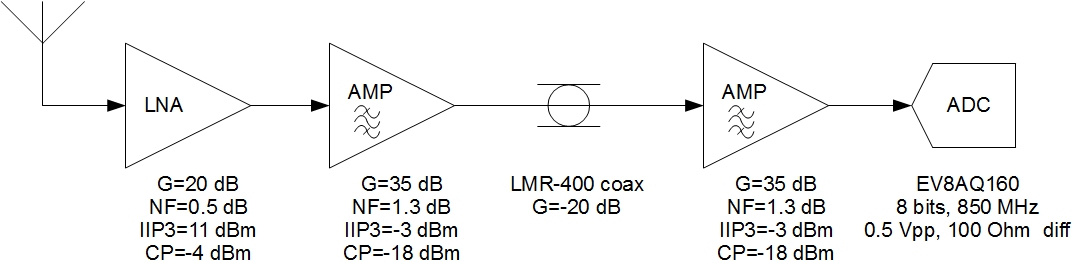}
    \caption{Coax-based receiver block diagram for one channel of the two-element 
      interferometer. \MDnewtext{The
      gain (G), noise figure (NF), input third order intercept point 
      (IIP3), and 1~dB input compression point (CP) shown for each block are referred to 
      the block's input.} }
    \label{chimerx_coax} \end{center} \end{figure}

A block diagram of the (coax-based) receiver for the CHIME two-element
interferometer is shown in Figure~\ref{chimerx_coax}. The
interferometer consists of two 9~m diameter parabolic dish
reflectors spaced 19~m apart on an east-west line. The feed for each dish is a
four-square above a ground-plane 
\JMnewtext{and is based on the design presented in \cite{ml}}.\margincomment{add citation to lueng $\checkmark$}
 The amplifier chain consists of a
custom low noise amplifier (LNA) followed by a second-stage amplifier
and filter. The feed analog stages are in individually shielded boxes.
A 50~m coaxial cable is connected to the focus receiver to transmit
the signal to an electronics hut containing the digitizer and
additional back end electronics. The 20~dB loss from the coax in
Figure~\ref{chimerx_coax} includes the loss in both the cable and
attenuators installed at each end of the cable. Inside the electronics hut, the
signal passes through a second combined amplifier and filter box and
is then fed into the digitizer and correlator. For a more detailed
description of the CHIME two-element interferometer refer to
\cite{gdp}.

The frequency response of the analog receiver is defined by a
commercial band-pass filter with a 3~dB passband in the approximate
range 475-800~MHz. The noise of the LNA is approximately $35~$K
($\mathrm{NF}=0.5$~dB)
%% I don't think we need to say "not including feed loss and ground spill", 
%% because you have stated this is the noise of the LNA, not the receiver noise.
across the CHIME band and this is the main
contribution to the noise of the receiver electronics. 
The output signals from the receiver are
directly sampled at 850~MSPS, recovering the second Nyquist zone 
from 425-850~MHz.\MDnewtext{There is no analog down-conversion stage.}The
 ADC samples are processed by a custom Field Programmable Gate Array (FPGA) based
correlator using an FX (Fourier transform followed by cross-multiplication) architecture: The signals are first put through
a poly-phase filter bank (PFB) and then Fourier transformed using a
2048-point Fast Fourier Transform. The data processing thereafter is separate for each frequency. The four signals
(two dishes and two polarizations) are cross-multiplied and averaged to complete the correlation.

\MDnewtext{The RFoF link, having relatively high input noise ($\mathrm{NF}\approx
27$ dB), is the last stage of the receiver chain with its output
connected directly to the ADC. A gain of at least 50 dB upstream of the
RFoF input is required to make its noise contribution negligible in
comparison to the front-end LNA.  To ensure the RFoF link has
negligible impact on the entire analog chain, it is required to have
\JMnewtext{an} output referred dynamic performance that is substantially better than
\JMnewtext{that of} the ADC.
The SFDR of the RFoF is determined by its IP3 and noise floor. This}
requirement translates to an output IP3 (OIP3) much larger than the
IIP3 of the ADC and an output noise floor much lower than
the input noise of the ADC\footnote{In general, the board where the
  ADC is mounted has some attenuation, mainly due to input matching
  and single-ended to differential input conversion for the ADC, so the
  input and output noise floor and IP3 of the ADC board will be different.}.

For the RFoF link, the IP3 determines the level of the third-order
intermodulation distortion products (IMD3) as a function of the output
signal level. In contrast, an ADC does not have a single, well defined
IP3. For a typical ADC, the IMD3 curve has three regions. For low
level input signals, the IMD3 products remain relatively constant
regardless of signal level. Near the ADC full-scale, the IMD3 products
behave as expected (3~dB increase per every dB increase in the input
power). The power level at which this begins depends on the specific ADC device.
Above the full-range, the ADC acts as an ideal limiter and the IMD3
products become large due to clipping. For a detailed
description of this subject see \cite{dch}.%Ref.~\cite{dch}.

We determine the IP3 of the ADC by
measuring the IMD3 products using two input tones such that the ADC
is close to full range (which is typically the IMD3 value reported on
ADC datasheets). A calculation of the ADC IP3 based on this
measurement yields an optimistic value, and an overestimation of the
ADC SFDR as long as the ADC is below full range\footnote{In this
  document SFDR always refers to the spurious free dynamic range as
  calculated from the two-tone third order intermodulation distortion
  products. This should not be confused with the SFDR specified in ADC
  datasheets which is obtained from a single-tone test as the ratio of
  the signal power to the power of the strongest spurious signal.}.
Thus, requiring that the OIP3 of the RFoF is much greater than this
measure of the ADC IIP3 is enough to guarantee that the distortion
generated by the RFoF is much smaller than that generated by the ADC.

\begin{figure} \begin{center}
    \includegraphics[width=0.6\textwidth]{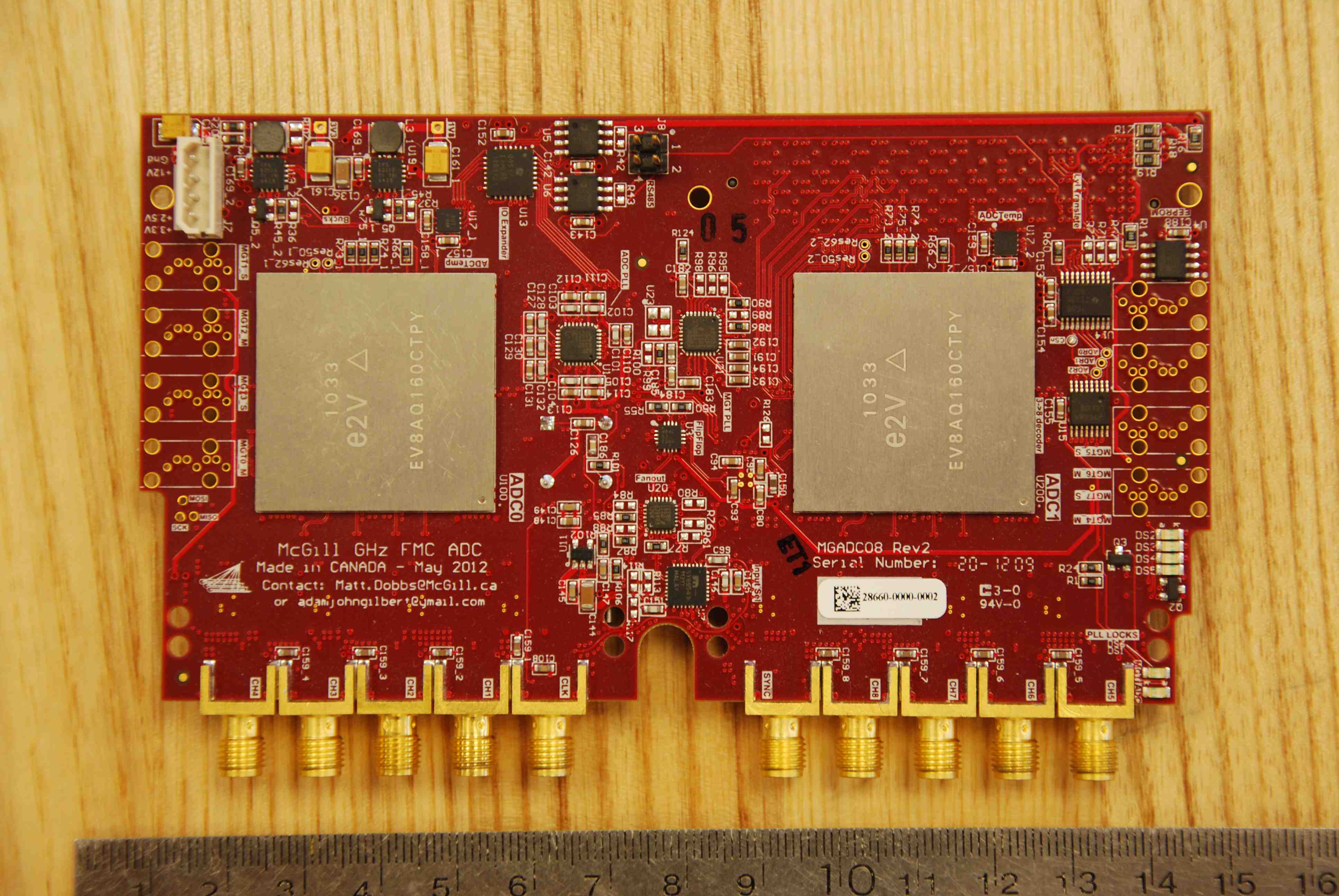} \caption{The custom-built ADC board for CHIME 
	includes two ADCs that can each digitize four analog channels with 8 bits at a maximum sampling
	rate of 1.25~GSPS, for a total of eight channels per board. Each
	ADC channel streams its data across eight differential signals to
	the FPGA. For CHIME, the ADCs will be operated at 800~MSPS to
	perform alias sampling of the 400-800 MHz signal. The ADC
	board can interface to any FPGA motherboard that
	implements the full FMC High
	Pin Count standard. The two additional inputs are for 
	the reference clock and the synchronization signal.}
\label{adc_board}
\end{center}
\end{figure}

\MDnewtext{Both the two-element interferometer and the CHIME}pathfinder
\JMnewtext{(a prototype of CHIME consisting of two cylindrical telescopes with 256 total radio receivers
that is currently under construction at DRAO)}use the
\MDnewtext{E2V\footnote{\href{http://www.e2v.com/}{http://www.e2v.com/}} EV8AQ160 quad-channel, 8-bit ADC
implemented on a custom FPGA Mezzanine Card (FMC) compliant circuit
board\margincomment{might want to add a photo of the ADC board to Fig
  5 $\checkmark$} with two devices providing eight input channels.
The insertion loss of the ADC board, from its single ended input to the differential
ADC, is approximately 2~dB.
Each ADC device
can sample at rates up to 1.25~GSPS in four-channel mode with a full range of
0.5~V$_\mathrm{p-p}$ differential input. The measurements presented in
this paper are sampled at 850~MSPS whereas CHIME will operate at 800~MSPS.} 
A top view of the custom ADC board is 
shown in Figure~\ref{adc_board}. \margincomment{Figures
2 and 3 on ADC measurements were removed. A photo of the ADC board was added instead $\checkmark$}
%Measurements of the ADC IMD3
%products in the CHIME band are shown in Figure~\ref{adc_imd3}. The
%IIP3 of the ADC is approximately 13~dBm. \margincomment{Should we remove figures
%2 and 3 on ADC measurements and instead include a photo of the ADC board?}

%\begin{figure} \begin{center}
 %   \includegraphics[width=1\textwidth]{adc_imd3} \caption{IMD3
 %     measurements are shown for input tones close to the ADC full scale using the
  %    custom ADC board %using a 2048-point FFT 
%      (left) and
 %     spectrum corresponding to the measurement at 7~dBFS (right). Two
 %     input tones at 632.5~MHz and 652~MHz were used. The ADC full
 %     range is 0.5~V$_\mathrm{p-p}$ or -5~dBm.}
%\label{adc_imd3}
%\end{center}
%\end{figure}

\JMnewtext{The performance of the ADC is characterized by the SFDR. 
The IP3 of the ADC board was measured using two tones at
632.5~MHz and 652~MHz. We obtained IIP3=13~dBm. The noise of the ADC board is determined from the
signal-to-noise-and-distortion (SINAD) measurement of the ADC
(typically parametrized as effective-number-of-bits, ENOB)
%which is parametrized as Effective-Number-of-Bits (ENOB) 
and then combined
with the IP3 measurement to obtain the SFDR. 
%Figure~\ref{adc_enob}
%shows the ENOB measurement at 632.5~MHz. 
The measured ENOB is 6.9~bits at 632.5~MHz. The corresponding input noise
floor is approximately $\mathrm{P}_{\mathrm{ni}}=-132$~dBm/Hz,
resulting in $\mathrm{SFDR}=97~\mathrm{dB}\cdot$Hz$^{2/3}$ for the
ADC board.}

%\begin{figure} \begin{center}
%    \includegraphics[width=0.8\textwidth]{adc_enob} \caption{ADC ENOB
%      measurement for the custom ADC board is shown.
%      % using a 2048-point FFT.
%      A 1 dBFS input tone at 632.5~MHz was used. The
%      first 25 folded harmonics have been identified where clearly
%      visible. The total power from noise and distortion in the
%      425-850~MHz band is approximately $-48$~dBm or
%      $\mathrm{SINAD}=43~$dBFS, which corresponds to
%      $\mathrm{ENOB}=6.9~$bits. The normalized output noise power is
%      $\mathrm{P}_{\mathrm{no}}=-134$~dBm/Hz.}
%\label{adc_enob}
%\end{center}
%\end{figure}

\section{RFoF link design and dynamic performance}
\label{RFoF_perf}

The selection of the components and modulation method for the RFoF
design is mainly driven by two parameters: cost and performance.
Due to its simplicity, intensity modulation combined with direct detection (IMDD)
is the most widely employed method to convey an RF signal over an
analog optical link \cite{aoltp} \cite{mpda} as compared to other modulation and detection techniques (e.g. frequency
and phase modulation of the optical carrier combined with coherent
detection)\margincomment{Added extra reference here $\checkmark$}. In 
IMDD the RF signal modulates the intensity of an optical carrier,
which then travels over the optical fiber and is detected by a
photodetector. Two ways to implement the IMDD method are to use
external modulation or direct modulation. In an externally modulated
system the laser operates in continuous wave mode and the
modulation is done externally with an optical modulator. These are
high performance systems that provide high dynamic range and large
bandwidth over long transmission distances (e.g. the implementation for ATA
\cite{ackcoxdre04}, \cite{welchetal09}). However, the need for
the additional external modulator and high performance, high power
lasers increases the system costs dramatically. This, in the case 
of a large-array application like CHIME requiring thousands of RFoF 
links, is a major disadvantage.

The custom-built RFoF link for CHIME is a directly modulated system.
%%MD below is repeated below, so I'm commenting out:
%where the laser current is directly modulated by the RF signal. 
The main advantage of direct modulation is simplicity and low
cost as compared to  externally modulated systems. For a short-range
application like CHIME, where the deterioration of the optical signal
in the fiber is small, the performance of the link is mainly
determined by the laser \cite{aoltp}.  \margincomment{Added reference here $\checkmark$}Distributed
 Feedback (DFB) lasers are
generally preferred for analog applications due to their high
linearity, high slope efficiency (the slope of the laser output power
versus the laser current curve) and low noise characteristic compared
to other lasers. However, these are the most expensive laser types due
to their complicated fabrication \cite{onpp}.
Vertical-Cavity-Surface-Emitting Lasers (VCSELs) offer low cost and
low power consumption due to the low threshold current. Their optical
characteristics are similar to DFB lasers (a single wavelength peak is
present in their spectrum) but their performance is, in general, below
that of DFB lasers. Several VCSELs were tested in the lab as possible
candidates for the RFoF prototype and SFDR values in the range 90-100
$\mathrm{dB}\cdot$Hz$^{2/3}$ were achieved. A similar RFoF link using
VCSELs for antenna remoting in the Australian Square Kilometre Array
Pathfinder achieved 98~$\mathrm{dB}\cdot$Hz$^{2/3}$
\cite{beresford08} \cite{deboeretal09}. These values are below the
requirements for CHIME. The current RFoF prototype uses a Fabry-Perot
(FP) laser, which for short-range applications offers good performance
and its costs are lower than for DFB lasers. Due to its
characteristic design, several wavelengths are present in the FP
spectrum, which results in signal deterioration for
transmission over \JMnewtext{distances exceeding a few kilometers}due
\margincomment{it'd be useful to say what you mean by ``long'' here, 10km? 1 km? $\checkmark$}to
dispersion in the fiber
\cite{aoltp} \cite{capmanyetal05} \cite{uhffo} \cite{wenworthetal92} 
\margincomment{Added extra reference here $\checkmark$}. For short-range applications like
CHIME, this effect is very small and the constraints from the laser
dominate. The performance of the link for distances larger than 100~m
has not been investigated.

\begin{figure} \begin{center}
    \includegraphics[width=1\textwidth]{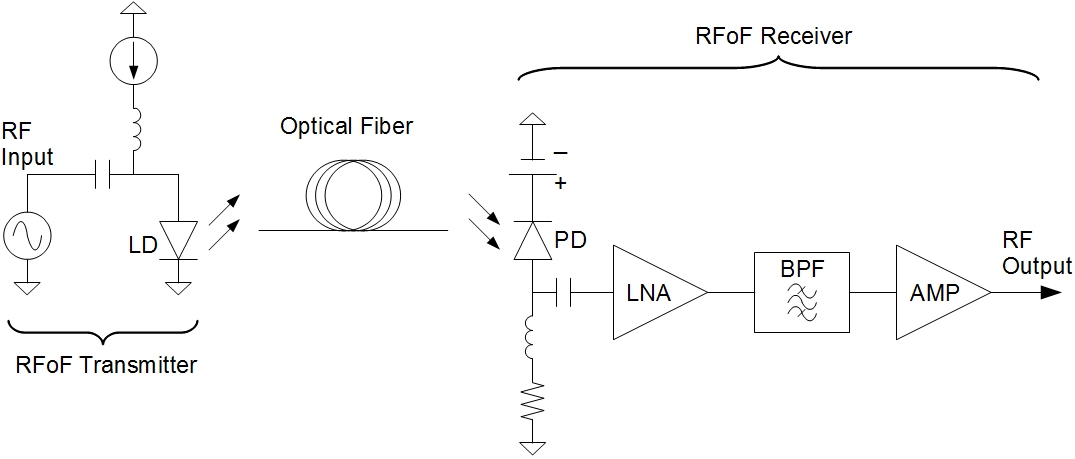} \caption{Schematic
      diagram of the prototype CHIME RFoF link. The laser (LD) current is
      directly modulated by the RF signal. The optical signal travels
      through the fiber and the photodiode (PD) converts the signal back to
      the electric domain. \JMnewtext{The RFoF receiver includes an
      amplification chain and the same band-pass filter (BPF) as the second
      stage amplifier of the two-element interferometer receiver.}}
\label{rfof_schem}
\end{center}
\end{figure}

Figure \ref{rfof_schem} shows the prototype RFoF link design. The RF analog
signal is placed on the optical carrier by directly modulating the laser
current with the signal. The optical signal travels through the
fiber and the photodiode converts the signal back to the electric
domain for additional amplification and filtering. The RFoF
transmitter features an
AGX\footnote{\href{http://www.agxtech.com/}{http://www.agxtech.com/}} 
uncooled, linear multi-quantum well (MQW) FP laser emitting at 1310
nm, hermetically sealed in an industry-standard coaxial package with a
single-mode fiber pigtail and subscriber connector/angled physical
contact (SC/APC) connector. Integrated in the laser package is an
optical isolator to prevent unwanted feedback into the laser and a
monitor photodiode to control the laser optical output power. Each
RFoF transmitter is electrically sealed in an aluminum box.

\begin{figure} \begin{center}
    \includegraphics[width=0.6\textwidth]{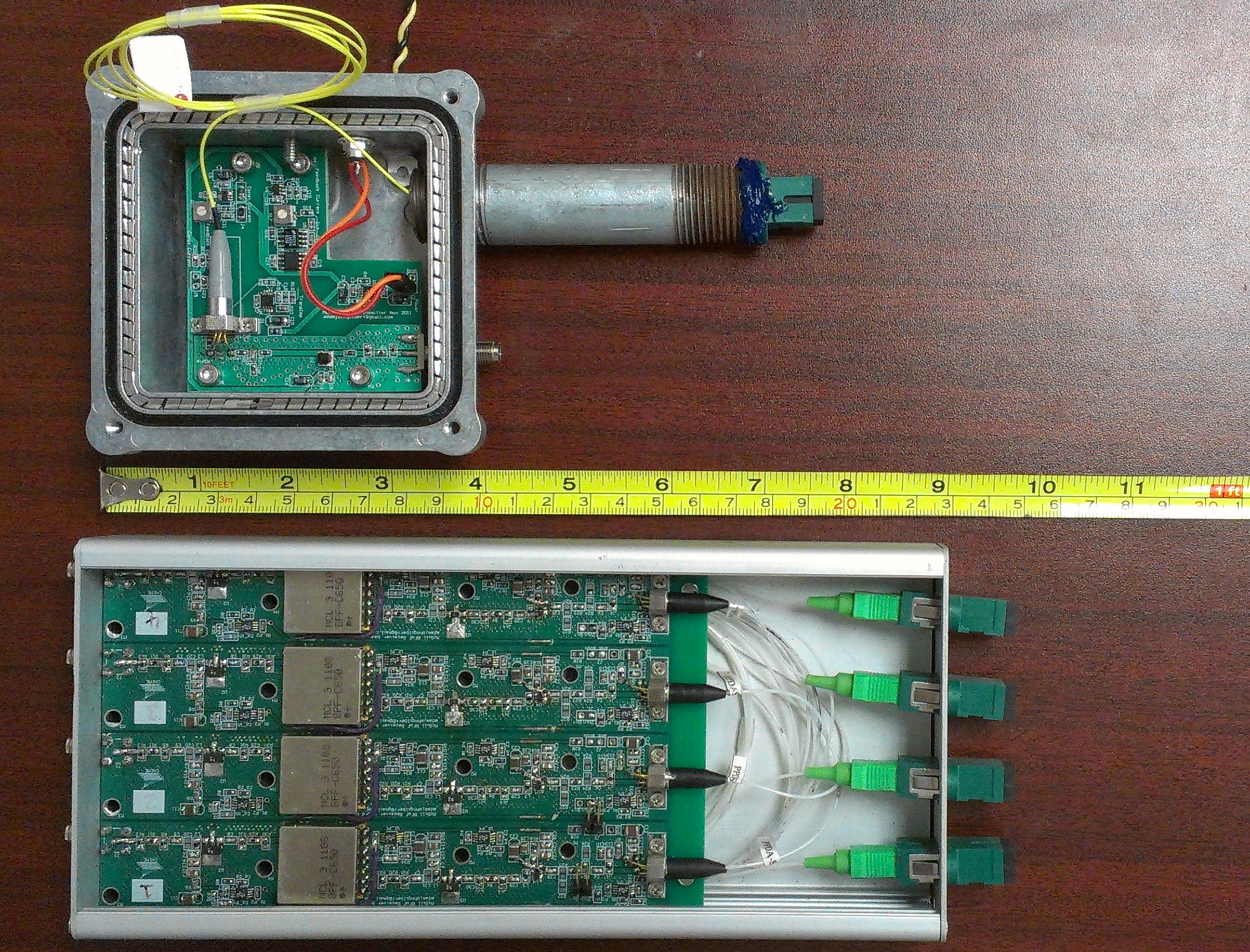} \caption{The RFoF
	transmitter (top) features an  
	uncooled, linear MQW FP laser emitting at 1310~nm, 
	hermetically sealed in an industry-standard coaxial package with a
	single-mode fiber pigtail and 
	SC/APC connector. Each
	RFoF transmitter is electrically sealed in a dedicated aluminum box.
	The RFoF receiver (bottom) uses a linear, low-capacitance photodiode sensitive
	to the wavelength range between 1100 nm and 1650 nm. This photodiode
	module is also a hermetically sealed coaxial package with a single-mode fiber pigtail and SC/APC connector.
	The first version of the RFoF receiver was
	constructed as a board with four receiver channels, all enclosed in
	the same aluminum box. 
	Each RFoF receiver channel includes an
	amplification chain and the same band defining filter as the second
	stage amplifier of the two-element interferometer receiver.}
\label{rfof_pic1}
\end{center}
\end{figure}

The RFoF receiver uses an AGX linear, low-capacitance photodiode that is sensitive
to the wavelength range between 1100 nm and 1650 nm. This photodiode
module is also a hermetically sealed coaxial package with a single-mode
 fiber pigtail and SC/APC connector. The\JMnewtext{prototype version of the} RFoF receiver was
constructed as a board with four receiver channels, all enclosed in
the same aluminum box.\margincomment{for CHIME, we have already moved
  to one receiver per box, due to cross talk. xtalk is nowhere
  mentioned in this paper. We should add a sentence or two about it,
  and warn that the new system is one ch per box. $\checkmark$}
Each RFoF receiver channel includes an
amplification chain and the same band defining filter as the second
stage amplifier of the two-element interferometer receiver in Figure~\ref{chimerx_coax}. A top view
of the single-channel RFoF transmitter and four-channel receiver is
shown in Figure \ref{rfof_pic1}. The
approximate cost is US\$~200 per link in quantities of several hundred,
not including the cost of fiber. The laser
and the band-pass filter dominate the cost.

\begin{figure} \begin{center}
    \includegraphics[width=1\textwidth]{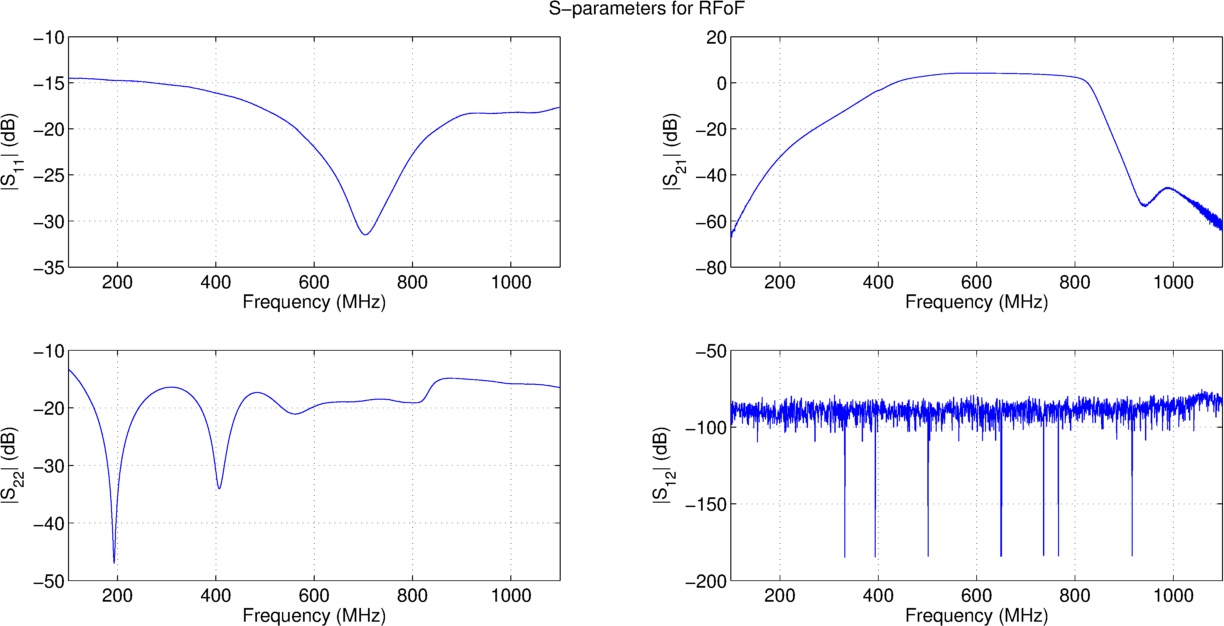}
    \caption{Scattering parameters of the RFoF link. An 80 m fiber was
      used. 
      S$_{11}$ (top left) and S$_{22}$ (bottom left) are, respectively, the input and output reflection coefficients of the link.
      S$_{21}$ (top right) and S$_{12}$ (bottom right) are, respectively, the forward and reverse gains.
      The average gain in the 475-800 MHz band is approximately
      4 dB. All measurements are performed at 25~$^\circ$C. The S$_{12}$
      measurement is limited by the dynamic range of the Network
      Analyzer used.}
\label{rfof_sparam}
\end{center}
\end{figure}

\begin{figure} \begin{center}
    \includegraphics[width=1\textwidth]{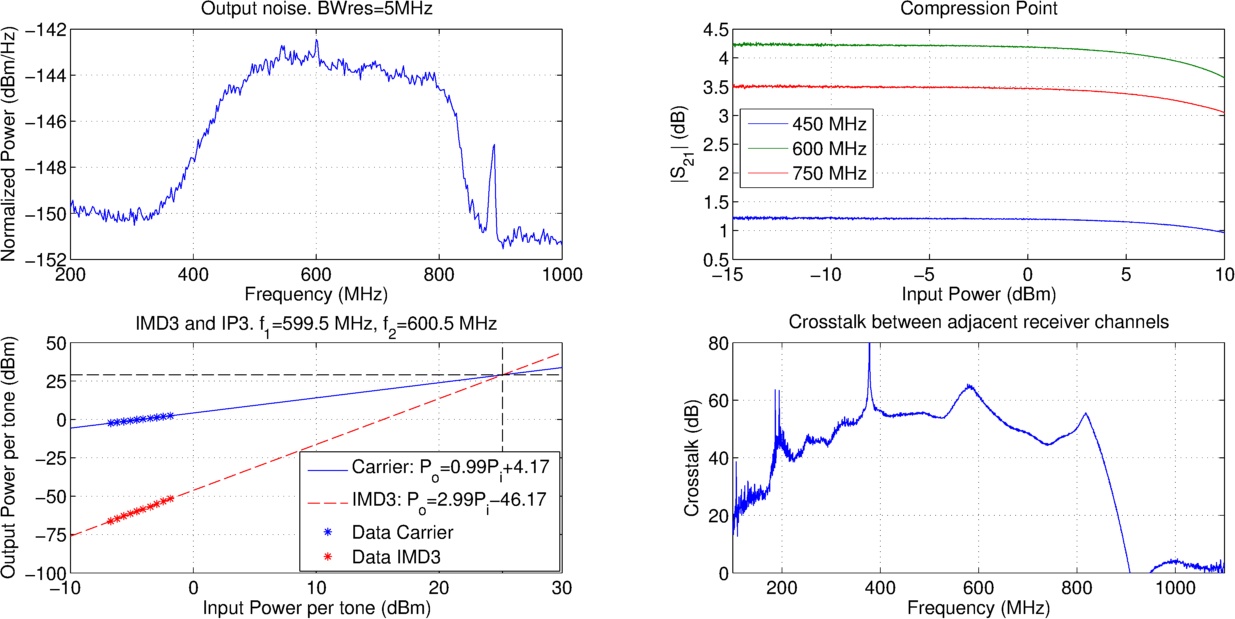}
    \caption{\JMnewtext{Output noise (top left), Compression Point (top right), IP3 (bottom left) and crosstalk between 
      adjacent receiver channels (bottom right) of the RFoF link
      at 25~$^\circ$C using an 80~m fiber. The average normalized output noise power in the
      475-800~MHz band (top left) is $\mathrm{P}_{\mathrm{no}}=-143$~dBm/Hz and
      the out-of-band noise floor is limited by the dynamic range of
      the Spectrum Analyzer. The CP exceeds 10~dBm
      and the measurement (top right) is limited by the maximum output power of
      the Network Analyzer.  The OIP3 is approximately 29~dBm (bottom left) measured using 
      two tones at 599.5~MHz and 600.5~MHz. The crosstalk performance between adjacent 
      RFoF receiver channels (bottom right) is  better than 45~dB across the CHIME band.}}
\label{rfof_No_CP_OIP3_xtalk}
\end{center}
\end{figure}

The performance of the RFoF was characterized with a suite of
laboratory tests. All the measurements in this section are performed
at 25~$^\circ$C using a 80~m fiber. The optical output power
feedback control circuitry of the RFoF transmitter has been disabled
for these tests as explained in Section~\ref{RFoF_stab}.
Without the RFoF receiver amplification chain, the gain of the link is
approximately -20 dB, a value which is mainly determined by the slope
efficiency of the laser and the responsivity of the photodiode
\cite{aoltp} \cite{coxackhel97}. Amplification is needed in the receiver to
boost the OIP3 of the RFoF to a level well above that of the ADC, while
keeping the output noise floor below the input noise of the ADC. This
is implemented in two stages. A commercial LNA is placed after
the photodiode, followed by the band defining filter and a
commercial high IP3 amplifier for the final amplification stage. An
amplification greater than the electric-optical-electric conversion
loss is required to have margin which allows for attenuation pads at
each stage, improving the impedance matching and reducing the in-band
ripple from the filter. The end-to-end gain of the RFoF link in the CHIME band
has been set to 4 dB with a corresponding OIP3 of approximately 29
dBm. Measurements of the scattering parameters for the RFoF link are
shown in Figure~\ref{rfof_sparam} and measurements of the output
noise, CP, OIP3 \JMnewtext{and crosstalk between adjacent RFoF receiver channels} are shown
in Figure~\ref{rfof_No_CP_OIP3_xtalk}. A summary of the characteristics and
dynamic performance of the RFoF link in the CHIME band is shown in
Table ~\ref{RFoF_perf_table}.

\begin{table}[h] \caption{\label{RFoF_perf_table}Characteristics and
    dynamic performance of the RFoF link in the 475-800 MHz band. All
    measurements are performed at 25~$^\circ$C using an 80 m fiber.}
\begin{center}
\begin{tabular}{|c|c|c|}
\hline

%Fiber Type & $9/125\mathrm{\mu m} $ single-mode \\ 
%& SC/APC connector \\\hline
\bf{Parameter} & \bf{Symbol} & \bf{Value} \\\hline

Operating wavelength& $\lambda$& $1310~\mathrm{nm} $ \\ \hline

Laser operating bias current & I$_{\mathrm{bias}}$ &  $30~\mathrm{mA} $ \\ \hline

Laser optical output power& P$_{\mathrm{opt}}$& $3.0~\mathrm{mW} $ \\ \hline

1 dB (input) compression point& CP&$>10~\mathrm{dBm} $ \\ \hline

Gain & G&$4~\mathrm{dB} $ \\ \hline

Input return loss &$|\mathrm{S}_{11~\mathrm{dB}}|$& $>16~\mathrm{dB} $ \\ \hline

Output third order intercept&OIP3& $29~\mathrm{dBm}$ \\ \hline

Output noise floor&P$_{\mathrm{no}}$& $-143~\mathrm{dBm/Hz} $ \\ \hline

Noise figure&NF &27 dB  \\ \hline

Spurious-free dynamic range&SFDR& $115~\mathrm{dB\cdot Hz^{2/3}} $ \\ \hline

\end{tabular}
\end{center}
\end{table}

%The noise of the RFoF link is dominated by the Relative Intensity
%Noise (RIN) of the laser, with an additional contribution due to Mode
%Partition Noise (MPN), a form of noise characteristic of Fabry-Perot
%lasers that arises from the random distribution of the photons
%produced by stimulated emission in each mode of the laser
%spectrum \cite{slf}. 
\JMnewtext{The main contribution to the noise of the RFoF link is the Relative Intensity
Noise (RIN) of the laser. Since the dispersion effects introduced by the fiber are small near 1310 nm and for the 
transmission distances considered in CHIME (approximately 100~m), the noise penalty 
introduced by the spectral impurity of the FP laser (e.g. Mode Partition Noise, a form of noise
that arises from the random distribution of the photons produced by stimulated
emission in each wavelength of the laser spectrum) is negligible \cite{wenworthetal92} \cite{slf}.}The
 RFoF link has
$\mathrm{NF}=27~$dB (P$_{\mathrm{ni}}=-147~\mathrm{dBm/Hz}$) and the
corresponding SFDR is
$115~\mathrm{dB\cdot Hz^{2/3}}$, or 57~dB in a 425~MHz
bandwidth. These results are well matched to an 8-bit ADC, as they
will have a negligible impact on the noise and linearity of
the receiver chain. This %RFoF system
performance over 80~m distances is comparable to \JMnewtext{that of} more
expensive high performance DFB laser-based RFoF links for radio
astronomy applications (e.g. \cite{monteetal09}, \cite{monteetal05},
\cite{perini09}).

The degradation of the ADC's SFDR introduced by the RFoF link can be
estimated by calculating the noise and IP3 of the cascaded RFoF+ADC
system at the input of the RFoF using the method of \cite{rfmta}. The
input noise of the cascaded system is
\begin{eqnarray}
\mathrm{P}_{\mathrm{ni}}=10\mathrm{log}\left(10^{\mathrm{P}_{\mathrm{no~RFoF}}/10}+
  10^{\mathrm{P}_{\mathrm{ni~ADC}}/10}\right)-\mathrm{G}_{\mathrm{RFoF}}
 \approx-135.7~\mathrm{dBm/Hz} . 
\end{eqnarray}
where $\mathrm{P}_{\mathrm{no~RFoF}}$ and $\mathrm{P}_{\mathrm{ni~ADC}}$ are the
output noise of the RFoF link and input noise of the ADC board (in dBm/Hz), respectively, and
$\mathrm{G}_{\mathrm{RFoF}}$ is the gain of the RFoF link (in dB).
The IIP3 of the cascaded system is
\begin{eqnarray}
\mathrm{IIP3}=-10\mathrm{log}\left(10^{-\mathrm{OIP3}_{\mathrm{RFoF}}/10}
  + 10^{-\mathrm{IIP3}_{\mathrm{ADC}}/10}\right)
-\mathrm{G}_{\mathrm{RFoF}} \approx8.9~\mathrm{dBm} .
\end{eqnarray}
where $\mathrm{OIP3}_{\mathrm{RFoF}}$ and $\mathrm{IIP3}_{\mathrm{ADC}}$ are the 
OIP3 of the RFoF link and IIP3 of the ADC board (in dBm), respectively.
Thus, the SFDR of the cascaded system is
$\mathrm{SFDR}\approx96.4~\mathrm{dB\cdot Hz^{2/3}}$, a reduction of
less than 1~dB with respect to the SFDR of the ADC alone.

\JMnewtext{The crosstalk performance between adjacent 
RFoF receiver channels is better than 45~dB in the CHIME band. The crosstalk requirement for CHIME
is set by the ADC board, which has a minimum crosstalk performance better 55~dB between
adjacent channels in the CHIME band. In order to meet this requirement, the production version of the RFoF link for CHIME
will use an aluminum shielded box for each individual RFoF receiver channel.}

\section{Gain and phase stability of the RFoF link}
\label{RFoF_stab}

The gain
%%MD: this footnote doesn't make any sense to me... if you're
%%reporting both gain and phase, then the gain probably refers to the
%%magnitude, and the phase is separate. Neither is a complex number
%%and there are no complex numbers shown in your plots.
%\footnote{In the following discussion, fluctuations in the
%  amplitude of the complex gain of the system are referred to as gain
% fluctuations.} 
and phase stability of the analog receiver are
critical for radio interferometry. Gain
and phase fluctuations, resulting mainly from temperature effects in
the components of the receiver, generate errors in the visibility data
and loss in sensitivity~\cite{isra}. The receiver stability determines how
often the calibration procedure, such as the injection of a known
signal across the array, needs to be performed.
% and even if it is still unclear what are the gain and phase stability requirements for the science goals in CHIME, 
%The stability is required to be such that the calibration procedure can be performed on intervals of at least several minutes.

In the case of the CHIME two-element interferometer, \JMnewtext{the RFoF receiver
is located in a temperature controlled electronics hut with the back end electronics}. 
Thus it is expected that the gain
and phase stability of the RFoF will be driven by temperature
fluctuations of the transmitter and the fiber, which are in an outdoor
environment. Measurements as a function of temperature will be
presented separately for these two components below.

\begin{figure} \begin{center}
     \includegraphics[width=1\textwidth]{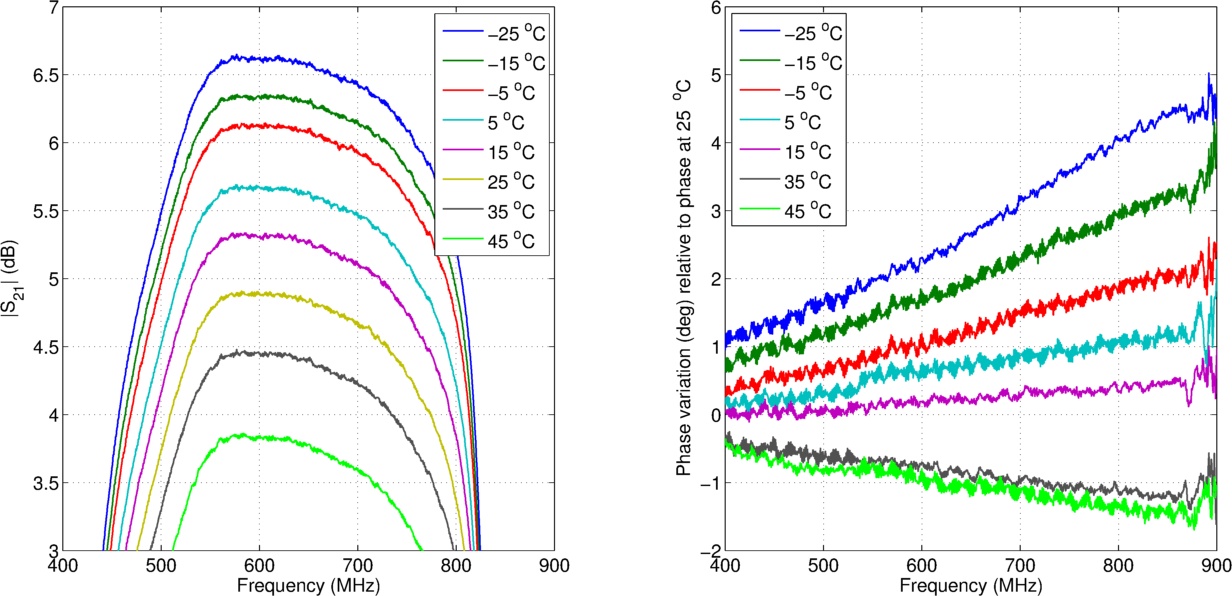}
     \caption{Gain (left) and phase (right) variation of the RFoF link with changes
       in temperature of the RFoF transmitter. An 80 m fiber was used.
       The temperature of the RFoF receiver and fiber was kept at 25~$^\circ$C.
       The phase variation is shown as the phase difference
       with respect to the phase at 25~$^\circ$C.}
\label{rfoftx_gain_phase_temp}
\end{center}
\end{figure}

Measurements of the gain and phase variation of the RFoF
link with changes in temperature of the RFoF transmitter box are shown
in Figure~\ref{rfoftx_gain_phase_temp}. The RFoF transmitter was
placed inside a climate controlled chamber and the temperature 
was varied from -25~$^\circ$C to 45~$^\circ$C in steps of
10~$^\circ$C. This encompasses the expected range of temperatures at
DRAO. Both the RFoF receiver and the 80~m fiber were kept at 25~$^\circ$C during the measurements. 
The phase variation is shown as
arg$(\mathrm{S}_{21}(T))-$arg$(\mathrm{S}_{21}(T=25~^\circ\mathrm{C}))$,
where arg$(\mathrm{S}_{21}(T=25~^\circ\mathrm{C}))$ is the
phase of $\mathrm{S}_{21}$ at 25~$^\circ$C and
arg$(\mathrm{S}_{21}(T))$ is the phase of $\mathrm{S}_{21}$
at the temperature of the chamber. The in-band gain variation is
approximately 3~dB for a 70~$^\circ$C temperature change, or
approximately 0.04~dB/$^\circ$C, mainly caused by the change in the
slope efficiency of the FP laser. The phase
variation is below 6$^\circ$ in-band for a 70~$^\circ$C temperature
change.

\MDnewtext{The RFoF transmitter includes an automatic optical power control
module for the laser. When the module is enabled, 
higher gain variations were observed.
This is due to}  a combination of a decrease in the slope efficiency of
the laser as the temperature increases and the fact that the
P$_{\mathrm{opt}}$~vs.~I$_{\mathrm{bias}}$ curve of the laser starts
to compress as the current is increased. 
Since the main purpose of the optical power control system is to keep
the laser bias current well above the threshold current of the laser
(which increases with temperature), it is possible to disable the
module as long as the laser bias current, which is now fixed, is well
above the threshold current along the full range of expected temperatures
for the RFoF transmitter. The power control module
has been disabled for the measurements presented in this paper and it will be
removed for the production version of the RFoF link for CHIME.

%\begin{comment}

\JMnewtext{In addition, the nonlinearity and noise performance of the RFoF link as a function of the 
 temperature of the RFoF transmitter were measured. 
Within the uncertainty of the measurements (about 1~dB),
the OIP3, output noise, and consequently the SFDR
remained constant during the test (at 29~dBm, $-143~\mathrm{dBm/Hz}$, and 
$115~\mathrm{dB\cdot Hz^{2/3}} $ respectively). 
However, the IIP3 and NF rose with temperature: % by about 3~dB from -25~C to 45~C: 
the IIP3 changed from 23~dBm at -25~$^\circ$C to 26~dBm at 45~$^\circ$C while
the NF changed from 25~dB at -25~$^\circ$C to 28~dB at 45~$^\circ$C.} \margincomment{JD: I removed the figure
with the OIP3 and outpurt noise as function of T. Just mention that we observed no change in OIP3 and Pno
within uncertainty.}

%\end{comment}

\begin{comment}

In addition, measurements of the behavior of the OIP3 and noise with
the temperature of the RFoF transmitter were performed. These are
shown in Figure~\ref{rfof_oip3_noise_temp}. The degradation in the
performance of the link is very small, less than 0.2~dB decrease in
the SFDR with respect to the value at 25~C.

\begin{figure} \begin{center}
    \includegraphics[width=0.8\textwidth]{rfof_oip3_noise_temp}
    \caption{OIP3 and output noise variation of the RFoF link with
      changes in temperature of the RFoF transmitter. An
      80 m fiber was used.}
\label{rfof_oip3_noise_temp}
\end{center}
\end{figure}

\end{comment}

\begin{figure} \begin{center}
    \includegraphics[width=1\textwidth]{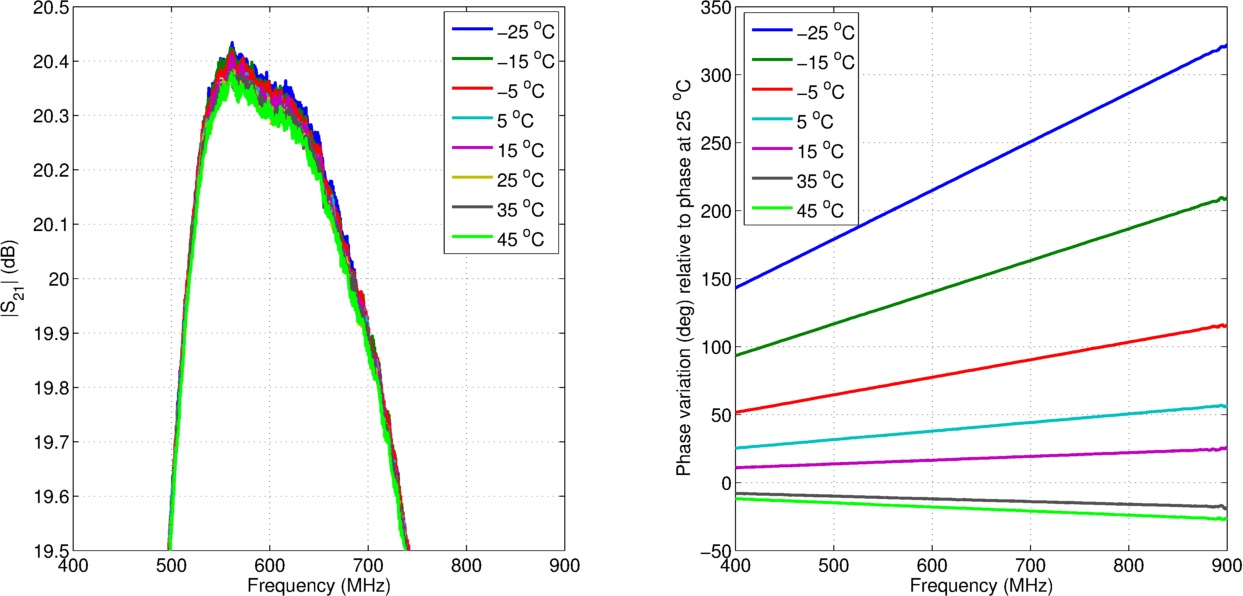}
    \caption{Gain (left) and phase (right) variation of the RFoF link with changes in
      temperature of the fiber. An 80 m fiber was used. The
      temperature of the RFoF transmitter and receiver was kept at 25~$^\circ$C.
      The phase variation is shown as the phase difference
      in degrees with respect to the phase at 25~$^\circ$C.}
\label{rfoffiber_gain_phase_temp}
\end{center}
\end{figure}

\begin{figure} \begin{center}
    \includegraphics[width=0.8\textwidth]{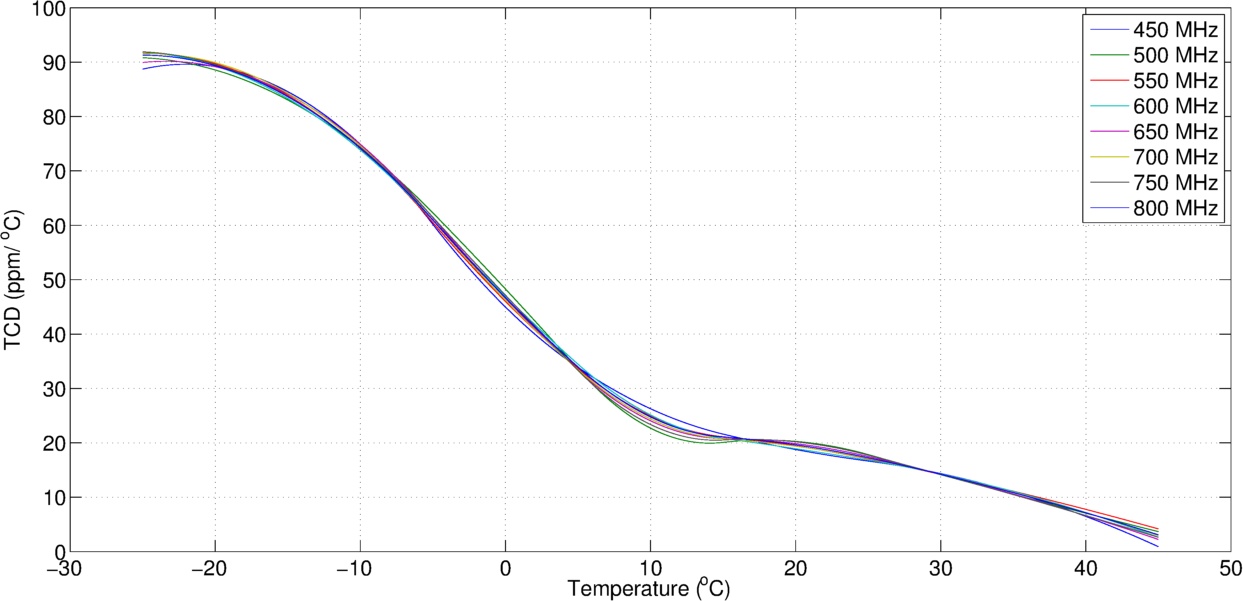}
    \caption{Estimated TCD of the fiber as a function of temperature. A velocity factor of 0.68 and
      a length of 80~m under standard conditions (25~$^\circ$C) was
      assumed for the fiber. Note that $\phi$ cannot be obtained
      directly from S$_\mathrm{21}$ since the insertion phase of the
      other components of the RFoF link has not been measured.}
\label{fiber_tcd}
\end{center}
\end{figure}

Measurements of the gain and phase variation of the RFoF link with
changes in temperature of the fiber are shown in
Figure~\ref{rfoffiber_gain_phase_temp}. The 80~m fiber was placed
inside the chamber and the temperature was varied from -25~$^\circ$C to 
45~$^\circ$C in steps of 10~$^\circ$C. A standard
9/125~$\mathrm{\mu}$m tight buffer patch cord was used. The gain
variation is less than 0.1~dB for a 70~$^\circ$C temperature change,
very small compared to the gain variation due to the RFoF transmitter.
However, the phase variation of the fiber is relatively high
and will be the dominant source of phase variations for the
interferometer. It is common to express the phase stability
of the fiber in terms of the Thermal Coefficient of Delay (TCD)
defined as
\begin{eqnarray}
\mathrm{TCD}=\frac{\mathrm{d}\tau}{\mathrm{d}T}\frac{10^6}{\tau}\approx\frac{\Delta \phi}{\phi}\frac{10^6}{\Delta T}
\end{eqnarray}
where $\tau$ is the insertion delay and $\phi$ is the
respective insertion phase of the fiber. $\mathrm{d}\tau/\mathrm{d}T$
is determined by variations in the length and refractive index of the
fiber~\cite{bergjohn83}. Figure~\ref{fiber_tcd} shows the estimated
TCD obtained from the phase information in
Figure~\ref{rfoffiber_gain_phase_temp} assuming that the fiber has a
velocity factor of 0.68 and a length of 80~m under standard conditions
(25~$^\circ$C). The TCD is approximately 90~ppm/$^\circ$C at -25~$^\circ$C,
 relatively high compared to LMR-400 coax cable, for which cable
providers typically specify a TCD below 10~ppm/$^\circ$C.
However, the phase stability of the optical link can be greatly
improved by using loose tube fiber, in which the manufacturing process
ensures that the fiber is protected from stresses caused by
temperature changes and mechanical forces \cite{ffotss}. For this
cable design the TCD reduces to values below 10 ppm/$^\circ$C
\cite{bergjohn83} and, in general, has a better performance in outdoor
applications. Since the change to loose tube causes no increase in the
cost of the cable, \MDnewtext{this will be the fiber used for the production
version of the CHIME RFoF link.}

\margincomment{MD: I'm on the fence as to whether the rest of this
  section should be kept for the paper.}
A method commonly used for the reduction of gain fluctuations
in the optical link is to control the temperature of the
laser by means of a thermistor and a thermoelectric cooler, with
both components either in a separate module on the RFoF transmitter
board or included in the laser package. The latter option is discarded
for CHIME because cooled lasers are typically an order of magnitude
more expensive than uncooled ones. Instead of trying to reduce the
gain fluctuations by analog means, a much simpler and more cost-effective
alternative is to perform a gain compensation of the analog receiver
implemented in firmware or software after the digitization or during the calibration
procedure.\MDnewtext{This is feasible as long as the signal degradation
introduced by ADC quantization and noise is acceptable across the full
range of temperature induced gain variations of the analog receiver.}

\begin{figure} \begin{center}
    \includegraphics[width=1\textwidth]{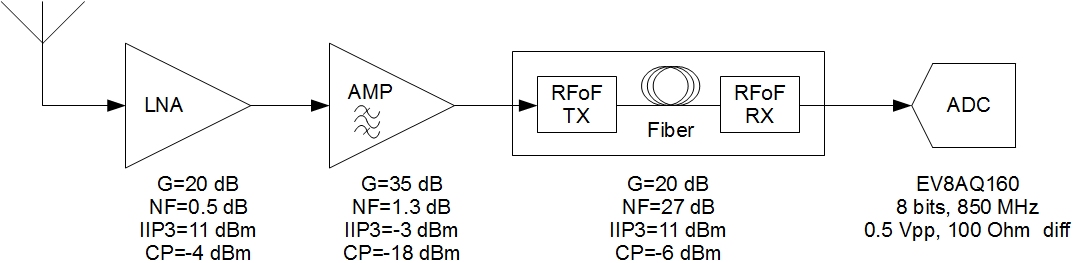}
    \caption{Two-element interferometer receiver block diagram using
      RFoF. The parameters of each block are referred to its
      respective input.}
\label{chimerx_rfof}
\end{center}
\end{figure}

\begin{figure} \begin{center}
    \includegraphics[width=1\textwidth]{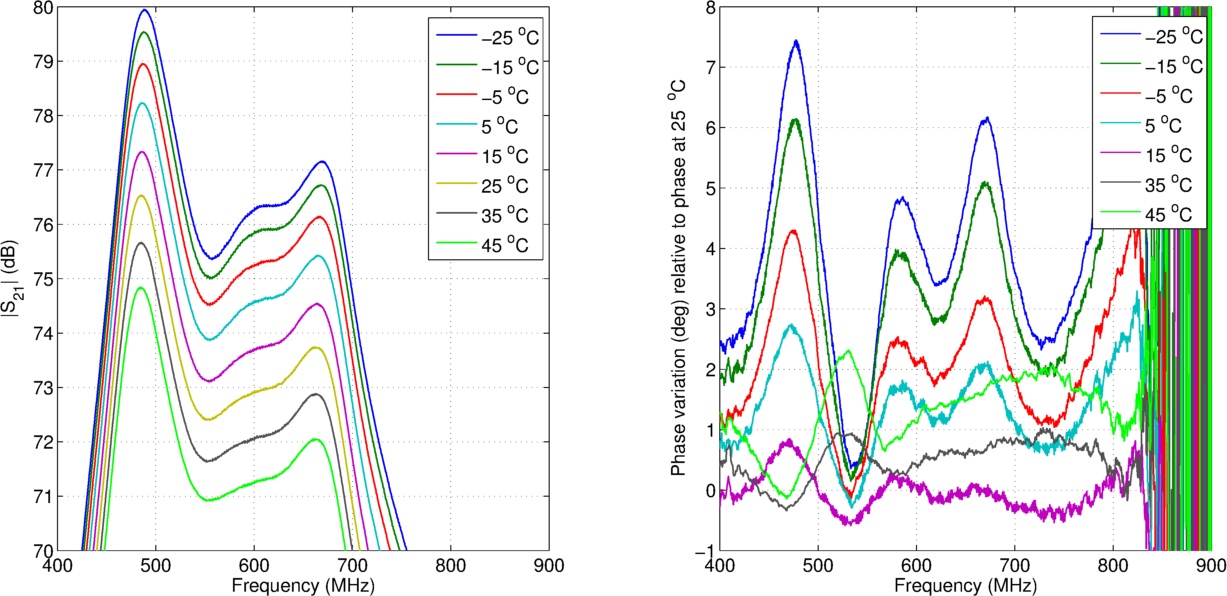}
    \caption{Gain (left) and phase (right) variation of the RFoF-based two-element
      interferometer receiver with changes in temperature of the
      elements at the focus. An 80 m fiber was used. The temperature
      of the RFoF receiver and fiber was kept at 25~$^\circ$C. The
      phase variation is shown as the phase difference with
      respect to the phase at 25~$^\circ$C.}
\label{chimerx_gain_phase_temp}
\end{center}
\end{figure}

In order to characterize this effect, the gain variation of the analog
receiver with RFoF was also investigated. A diagram of the receiver
chain with RFoF is shown in Figure~\ref{chimerx_rfof}. 
\JMnewtext{The current analog receiver design requires a 20~dB gain RFoF link 
in order to achieve the desired noise power at the input of the ADC. Optimally, the additional gain must be
added at the input of the RFoF link in order to keep the RFoF output noise level below that of the ADC.
For this test, the additional gain was added in the RFoF receiver so we could use the existing analog
receiver electronics. The effect of this change in the overall system performance is negligible since the noise
performance of the system is still dominated by the LNA and its nonlinear behavior is still dominated by the ADC.
The production version of the RFoF link will have the additional gain in the transmitter.}For this test,
%An additional
%20~dB amplification stage
\margincomment{MD: we need to explain this better, why didn't be put
  the gain upstream of the RFoF, etc. $\checkmark$}
% was added to the RFoF receiver in order to
%achieve the desired noise power at the input of the ADC. This may not
%be the optimal solution in terms of the SFDR of the ADC but allowed us
%to use the setup and components of the coax-based receiver. 
all the components of the receiver which are located at the
focus (LNA, second-stage amplifier and RFoF transmitter) were placed
in the chamber and the temperature inside was varied from
-25~$^\circ$C to 45~$^\circ$C. The results are shown in
Figure~\ref{chimerx_gain_phase_temp}. The in-band gain variation is
approximately 5~dB for a 70~$^\circ$C temperature change, or
approximately 0.07~dB/$^\circ$C.

\begin{figure} \begin{center}
    \includegraphics[width=0.8\textwidth]{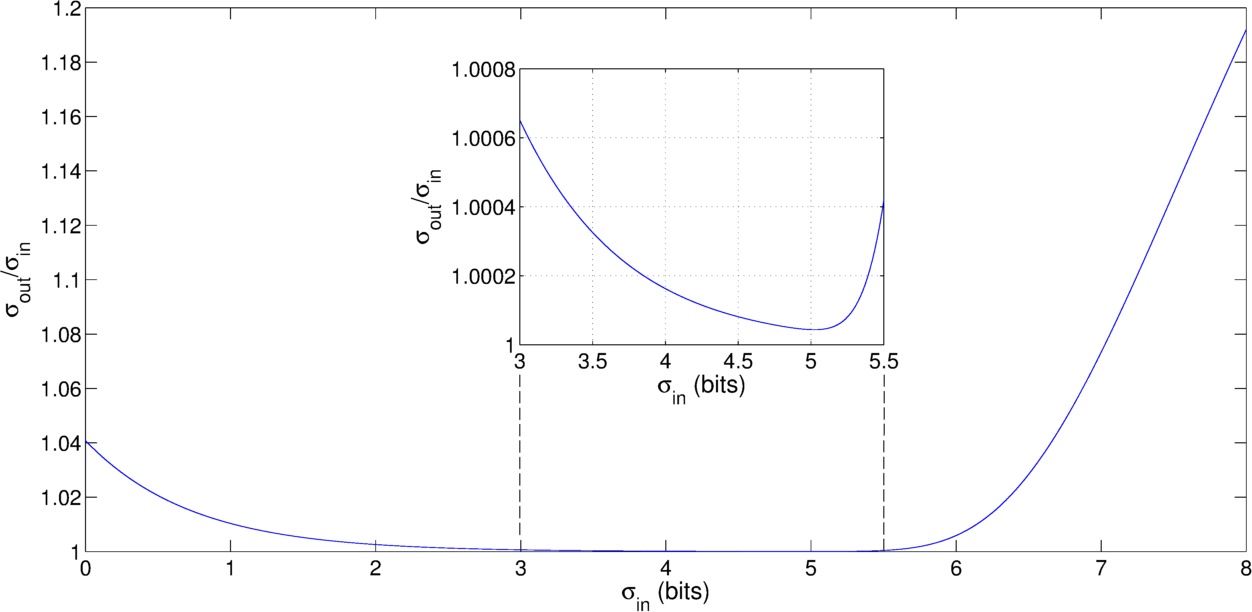}
    \caption{Fractional increase in the noise of a signal that results
      from the quantization with an 8-bit ADC. The input to the ADC is
      assumed to be gaussian noise with standard deviation
      $\sigma_{\mathrm{in}}$. The standard deviation of the signal
      after quantization is $\sigma_{\mathrm{out}}$.}
\label{quant_eff}
\end{center}
\end{figure}

The signal degradation due to quantization and 8-bit ADC noise 
is shown in Figure~\ref{quant_eff}. For a detailed
discussion of the fractional increase in the noise of a Gaussian
distributed input signal that results from quantization refer to
\cite{isra}. For the two-element interferometer, the typical input
power to each channel of the analog receiver is estimated as
$100~$K in a 425~MHz bandwidth, or about -92~dBm \cite{gdp}.
The average gain of the
receiver in the 425-850~MHz band is approximately 72~dB at 25~$^\circ$C
which, \JMnewtext{including the} 2~dB insertion loss of the ADC board, results in
-22~dBm power at the input to the ADC or a voltage standard deviation
of about 3.7 bits. This level causes a negligible noise penalty due to
quantization and ADC noise, while allowing headroom for receiver
linearity and ADC saturation caused by external RFI. \JMnewtext{The total variation in the gain of the receiver (from
-25~$^\circ$C to 45~$^\circ$C)} results
in an input signal variation between 3.4 bits (at 45~$^\circ$C) and 4.2 bits (at -25~$^\circ$C). As
Figure~\ref{quant_eff} shows, \JMnewtext{the minimum of the quantization efficiency
function is broad and the noise penalty at these levels
remains negligible, so the gain calibration can be performed after
digitization.}

\section{RFoF Characterization on the CHIME two-element interferometer}
\label{RFoF_test}

\MDnewtext{The first tests of the RFoF link on the CHIME two-element interferometer
at DRAO were performed in January 2013. The purpose of these tests was
to compare the performance of the RFoF-based
receiver to that of the coax-based receiver.}
At the focus of each dish, the output of the feed is sent through the
LNA, second-stage amplifier, and filter box. It is then split with one 
output connected to the ADC
through the RFoF system (Figure~\ref{chimerx_rfof}) while the other output
connected to the ADC through coax cable and a second combined
amplifier and filter box located inside the hut
(Figure~\ref{chimerx_coax}).

\begin{figure} \begin{center}
    \includegraphics[width=1\textwidth]{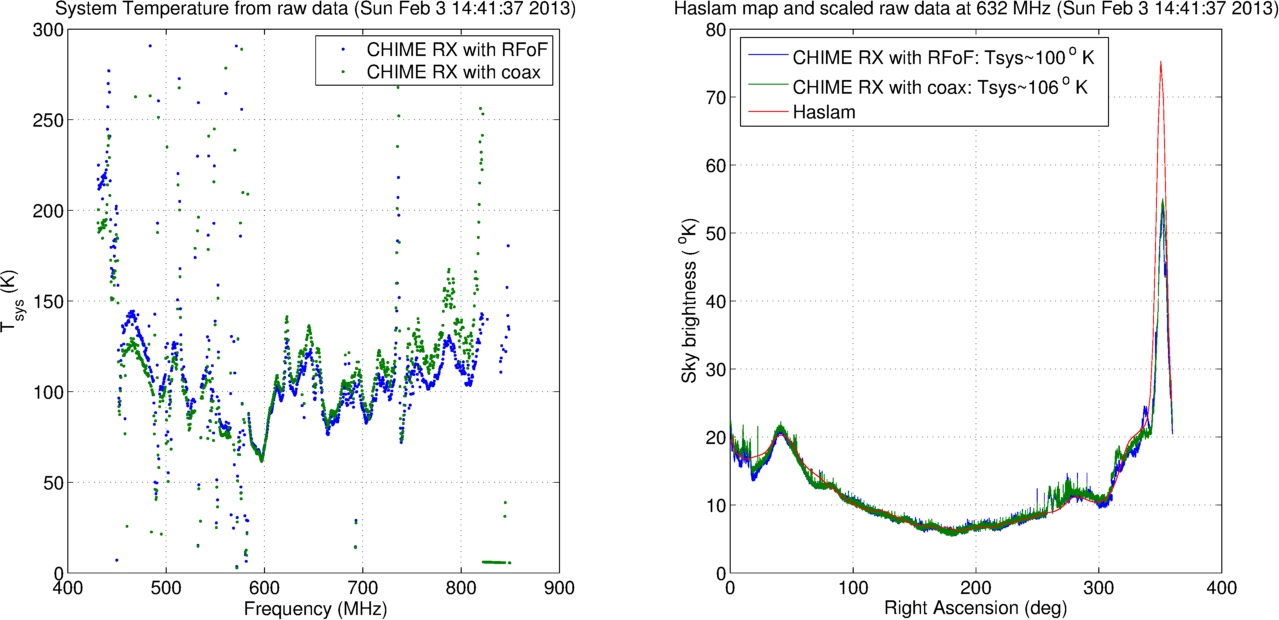}
    \caption{Estimated system temperature, $T_{sys}$, for the
      coax-based and fiber-based receiver across the CHIME band from
      raw data measured by comparing to the Haslam map excluding Cas A (left). The
      Haslam map was convolved with a Gaussian kernel determined by
      the beam parameters of the dish, and scaled using an estimated
      spectral index $\alpha_{est}=-2.5$.  
      A particular fit at 632 MHz
      is shown in the right plot.}
\label{drao_haslam}
\end{center}
\end{figure}

With this setup, we proceeded to compare the noise performance
of the two systems. \MDnewtext{The system temperature was calculated from 24
hours of raw data recorded in drift-scan mode and comparing it to the
408 MHz Haslam map \cite{haslam82}, excluding the bright radio source
Cas~A.} The Haslam map was convolved with a Gaussian kernel determined
by the beam parameters of the dish, and scaled using an estimated
spectral index $\alpha_{est}=-2.5$ (see \cite{gdp} for details). As Figure \ref{drao_haslam} shows,
the noise performance of the coax-based and fiber-based receiver
\MDnewtext{are consistent to within about 5~K across the band, which falls within the
systematic uncertainty of this measurement.}
The system temperature of both systems is roughly $T_{sys}\approx
100~$K midband, of which about $40~$K is caused by ground
spill and loss in the feed. 
\JMnewtext{The spectral ripple in the $T_{sys}$ measurement is caused by standing 
waves between the ground plane and the reflector.} 
\margincomment{MD: I don't think we can include the rest of this
  section unless we clearly understand why the correlation ceff
  differs from 1 $\checkmark$.}

%The fiber-based
%receiver has a slightly better performance at high frequencies, but
%this is hard to quantify since this calculation suffers from a large
%uncertainty caused by gain fluctuations and uncertainties in the beam
%parameters.

\section{Future work and conclusions}
\label{RFoF_conc}

In this paper, a prototype version of a low cost RFoF link for CHIME
has been presented. The RFoF was characterized with a suite of
laboratory tests and a very good dynamic performance was achieved,
comparable to more expensive high performance DFB laser-based
RFoF links used elsewhere for radio astronomy applications. In addition, the
gain and phase stability of the RFoF link were investigated. It was
shown that the gain fluctuations for the fiber-based CHIME receiver
cause only a small noise penalty due to quantization and ADC noise of
an 8-bit ADC, so the gain calibration can be performed after
digitization. Finally, it was shown that
% the noise performance of the fiber-based receiver is as good as that of 
\JMnewtext{there is no measurable noise degradation as compared to} 
the coax-based
receiver. \JMnewtext{Near term} improvements to this RFoF system \JMnewtext{will} include separating the RFoF
receivers into individually shielded boxes to improve the crosstalk
\margincomment{MD: its funny to mention crosstalk here, not having
  presented a measurement in the text. better to move this statement
  to the text section $\checkmark$}
performance and using loose tube fiber 
%%MD: it's not clear that we will do this, so do not mention
% and appropriate fiber connectors (e.g. FC/APC) 
to improve the phase stability in an outdoor environment. Valuable experience was
obtained during the design and characterization of the RFoF link for
CHIME, demonstrating that RFoF can be successfully applied for analog
signal transport in large-array radio astronomy applications at low cost.

\acknowledgments

We thank the DRAO staff and members of the CHIME collaboration for their comments
and stimulating discussions. We acknowledge funding
from the Natural Sciences and Engineering Research Council of Canada, the Canadian
Institute for Advanced Research, and the Canadian Foundation for
Innovation. MD acknowledges support from the Canada Research Chairs 
program.\JMnewtext{JM acknowledges support from the Fonds de recherche du Qu\'{e}bec-Nature et technologies.}

\end{document}